\documentclass[a4paper, 10pt, conference]{ieeeconf}      % Use this line for a4 paper

\IEEEoverridecommandlockouts                              % This command is only needed if 
\usepackage{graphicx}
\title{\LARGE \bf
Guidance of the Center of Pressure Using Haptic Presentation
}
\author{Yohei Kawasaki$^{1}$ and Yuta Sugiura$^{2}$% <-this % stops a space
\thanks{$^{1}$Yohei Kawasaki is with the Department of Information and Computer Science, Keio University, Yokohama, Kanagawa, Japan
        {\tt\small yohei\_green@keio.jp}}%
\thanks{$^{2}$Yuta Sugiura is with the Department of Information and Computer Science, Keio University, Yokohama, Kanagawa, Japan
        {\tt\small sugiura@keio.jp}}%
}

\begin{document}

\maketitle
\thispagestyle{empty}
\pagestyle{empty}

%%%%%%%%%%%%%%%%%%%%%%%%%%%%%%%%%%%%%%%%%%%%%%%%%%%%%%%%%%%%%%%%%%%%%%%%%%%%%%%%
\begin{abstract}
Accurately instructing posture and the position of the body's center of gravity is challenging. In this study, we propose a system that utilizes haptic feedback to induce the Center of Pressure (CoP) movement. The Wii Balance Board is employed to sense the CoP, and vibration motors are used for haptic feedback. To provide a comparison, inductions were also performed using visual and auditory feedback, and the time required for induction was measured. Additionally, after the experiments, a questionnaire survey was conducted.

\end{abstract}

%%%%%%%%%%%%%%%%%%%%%%%%%%%%%%%%%%%%%%%%%%%%%%%%%%%%%%%%%%%%%%%%%%%%%%%%%%%%%%%%
\section{INTRODUCTION}

While it's possible to convey information about body movements and the positioning of one's center of gravity through words or illustrations, accurately comprehending and embodying these instructions is challenging. For instance, individuals engaged in sports might record their pitching or swinging forms on video to aid in improvement. However, it's believed that relying solely on visual feeedback has its limitations when it comes to refining form. Therefore, sensing human body movements and providing feedback in some form is essential.

Many studies have been conducted to guide human movement using methods other than visual feedback. Wilko et al. \cite{1} proposed a belt for assisting in navigation. This belt is equipped with vibration motors that provide haptic feedback to indicate the direction to proceed. Marco et al. \cite{2} employed a bracelet-like device for haptic feedback, suggesting directions in skiing scenarios and facilitating communication with instructors. Luces et al. \cite{3} also utilized a technique known as "Phantom Sensation" to provide haptic feedback on the wrist, enabling human movement within a space.

\begin{figure}[t]
\begin{center}
\includegraphics[width=8cm]{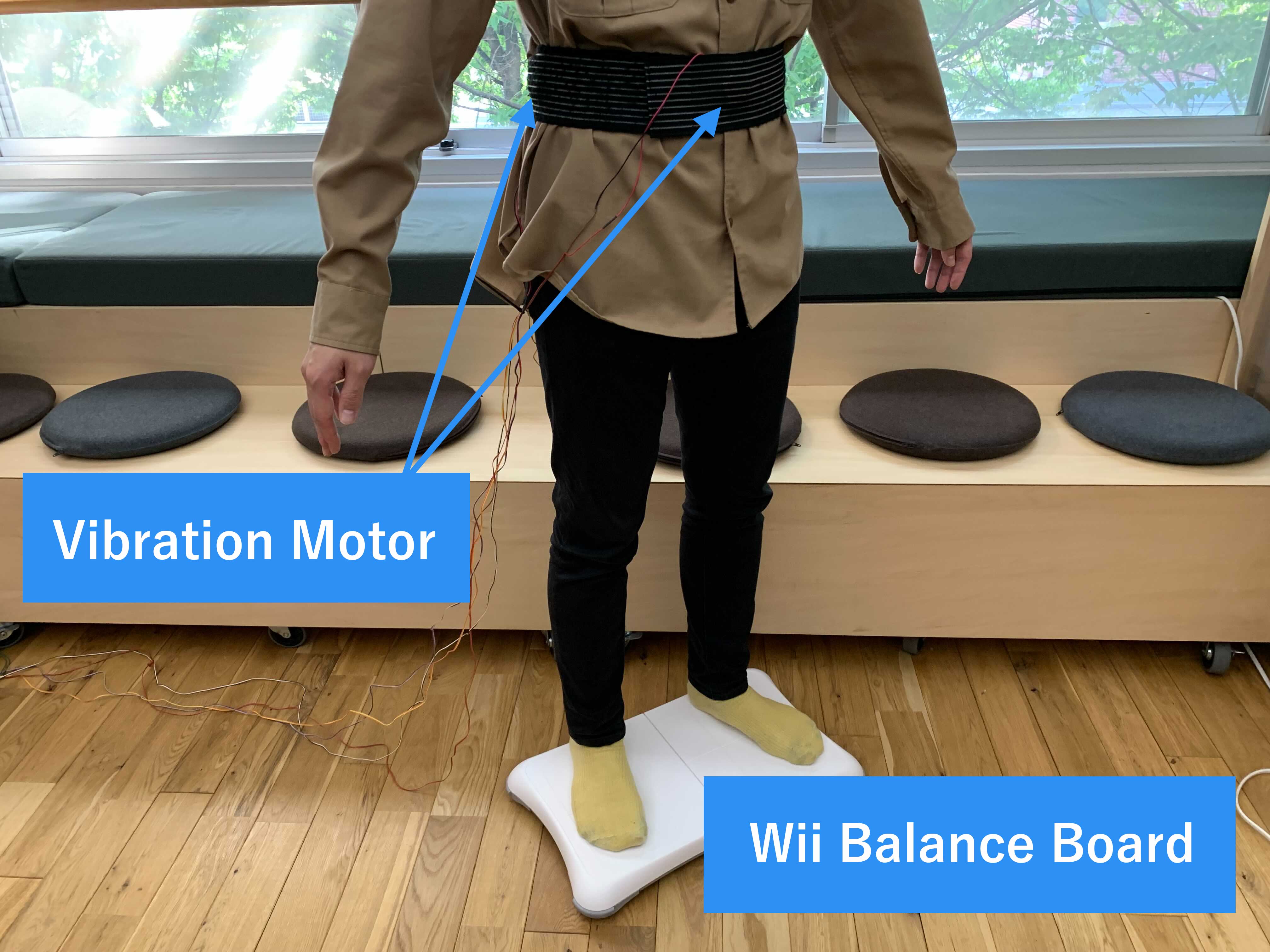}
\caption{\label{fig1} Wii Balance Board and Vibration Motor}
\end{center}
\end{figure}

In this paper, we propose a system aimed at instructing posture by using haptic feedback to guide the position of the CoP. The CoP sensing was performed using the Nintendo Wii Balance Board. Participants wore vibration motors around their waist in four directions: front, back, left, and right. They used haptic feedback to move the CoP accordingly (Fig.\ref{fig1}). For comparison, guidance through visual feedback and guidance through auditory feedback were also provided. Six participants were tasked with moving the CoP to a target location, and the time required for this task was measured. The results of a multiple comparison test showed no significant difference between guidance through haptic feedback and guidance through visual feedback.

\section{RELATED WORK}

\subsection{Guidance by haptic presentation}
There is a considerable body of research focused on guiding human movement using haptic feedback. The motivation behind providing haptic feedback extends beyond mere feedback; many studies also aim to facilitate communication with individuals who are visually impaired. Various parts of the body have been explored for delivering haptic feedback.
Wilko et al. \cite{1} proposed a navigation support belt that provides haptic feedback on the user's waist to assist with directional guidance. Studies have also investigated wrist-mounted bracelet-type devices for haptic feedback. For instance, Marco et al. \cite{2} suggested direction guidance for skiing and communication with instructors. Luces et al. \cite{3} employed the "Phantom Sensation" technique to facilitate human movement within a space. Huppert et al. \cite{4} devised a system where the movement of a drone is conveyed to the fingertips using strings, enabling subjects to identify object positions.
Elvitigala et al. \cite{5} embedded vibration motors in shoes to provide feedback for instructing posture during exercise.
In this study, we propose a system where participants wear vibration motors around their waist to guide the position of the CoP.

\subsection{Use of the Wii Balance Board}

\begin{figure}[t]
\begin{center}
\includegraphics[width=8cm]{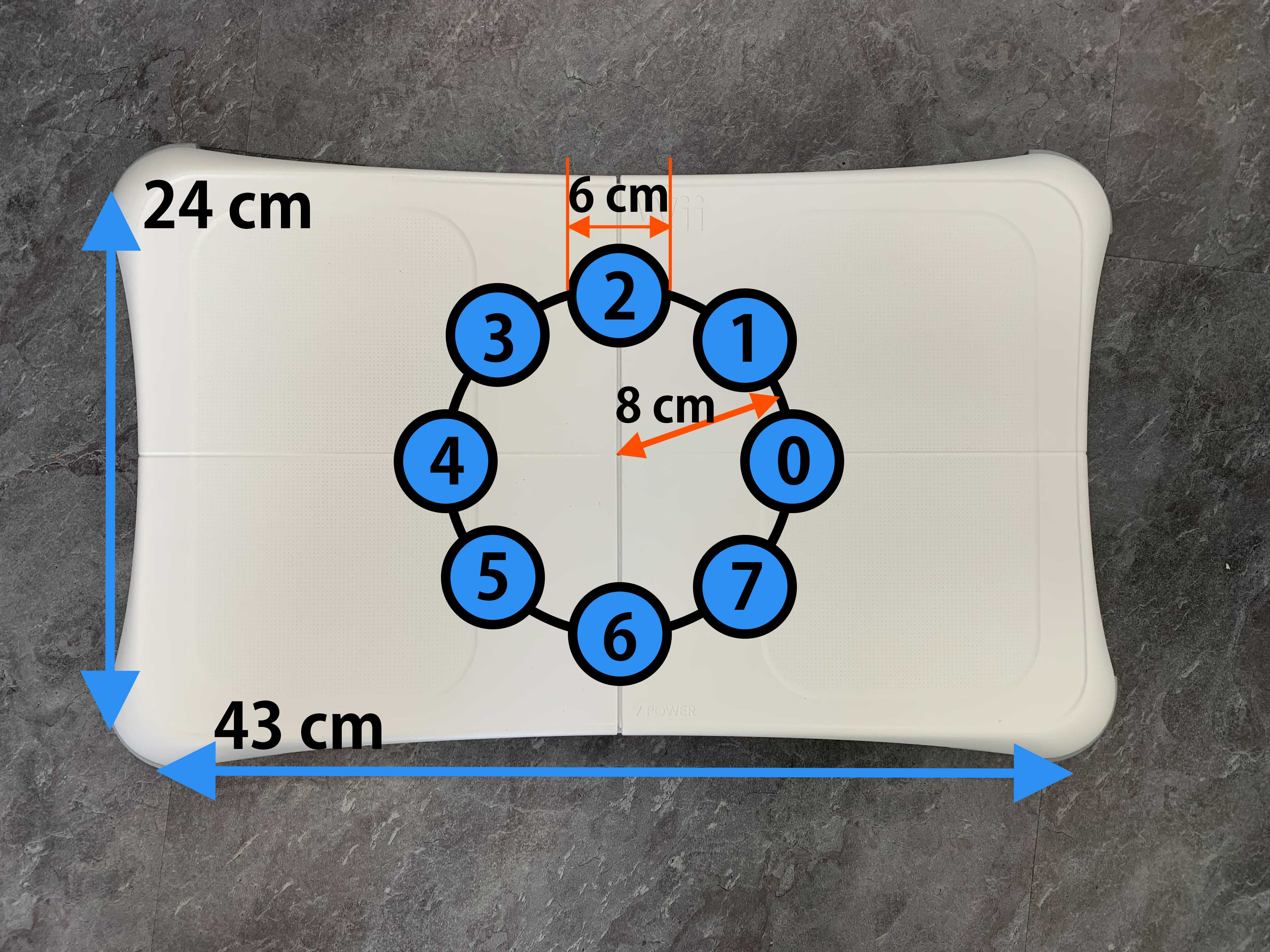}
\caption{\label{copmap} Wii Balance Board}
\end{center}
\end{figure}
CoP is the point where the forces acting on a human's feet converge, and it is frequently utilized for the purpose of posture control \cite{5}. While some methods employ pressure sensors for measuring CoP \cite{6}, there are also studies that use the Wii Balance Board. Clark et al. \cite{7} investigated the utility of the Wii Balance Board and concluded that, with proper data collection and processing methods, it could yield a similar level of reliability as force platforms for static standing computerized posturography.
Gil-Gómez et al. \cite{8} proposed a rehabilitation system to enhance upright balance in patients with ABI (Acquired Brain Injury). Holmes et al. \cite{9} used the Wii Balance Board for balance measurements in patients with Parkinson's disease.
Research involving the analysis of human posture using the Wii Balance Board has been conducted.
In this study, we measured CoP using the Wii Balance Board.
\subsection{Analysis of CoP}
Research involving the measurement and analysis of CoP is widely conducted. Farago et al. \cite{10} proposed a wearable device for evaluating walking physiology using CoP and EMG measurements. Ishida et al. \cite{11} suggested that changes in CoP position during bilateral leg squats primarily affect ankle and knee extension moments, and visual feedback of CoP could potentially assist in their correction. Based on these findings, it can be concluded that providing feedback and inducing changes in CoP position holds significance. In this study, we utilized the Wii Balance Board to measure CoP and employed haptic feedback to induce position changes.

\section{Experiment}\label{AA}

\begin{figure}[t]
\begin{center}
\includegraphics[width=7cm]{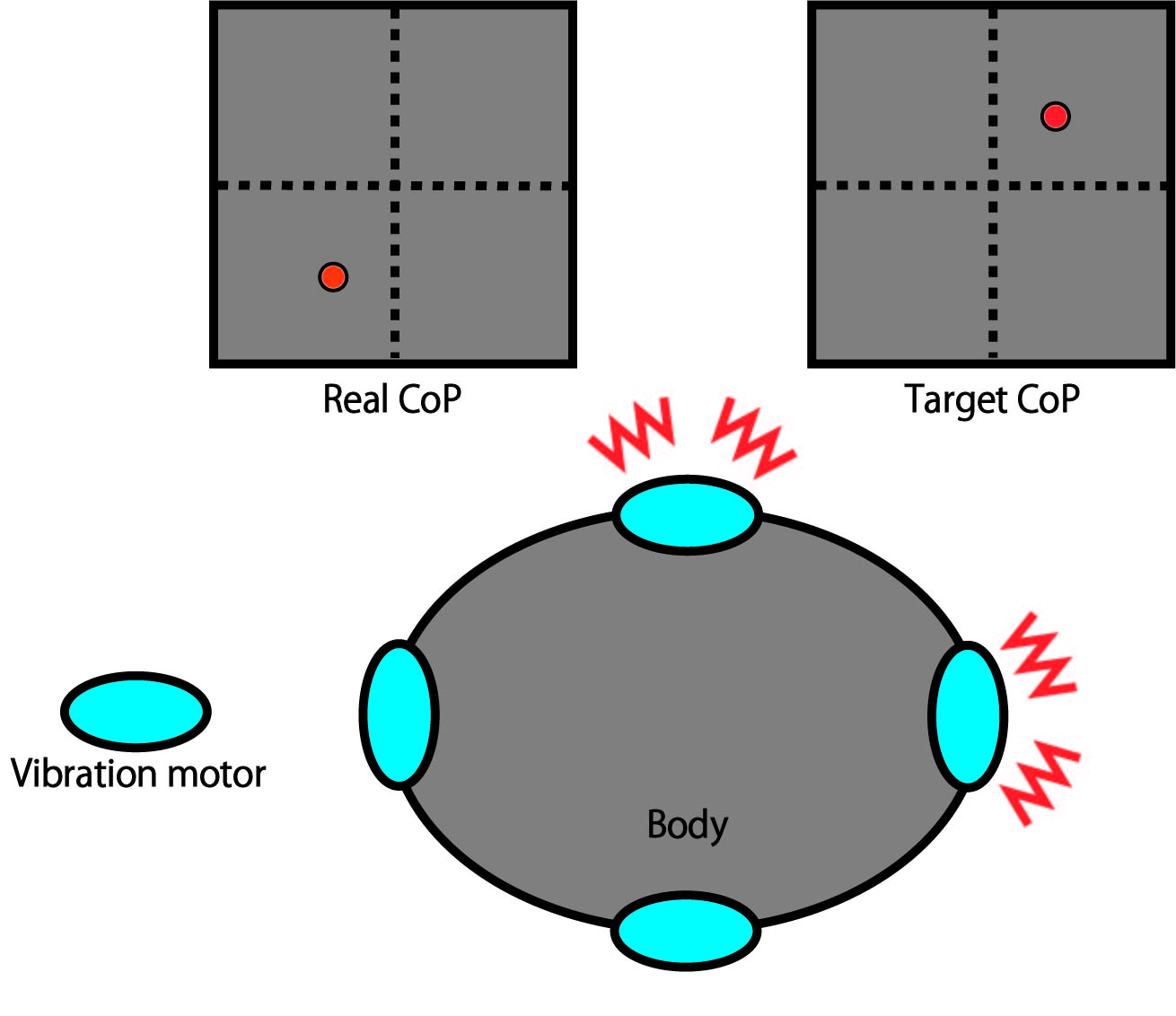}
\caption{\label{vib} How to Vibrate}
\end{center}
\end{figure}

\begin{figure*}[t]
\begin{center}
 \includegraphics[width=14cm]{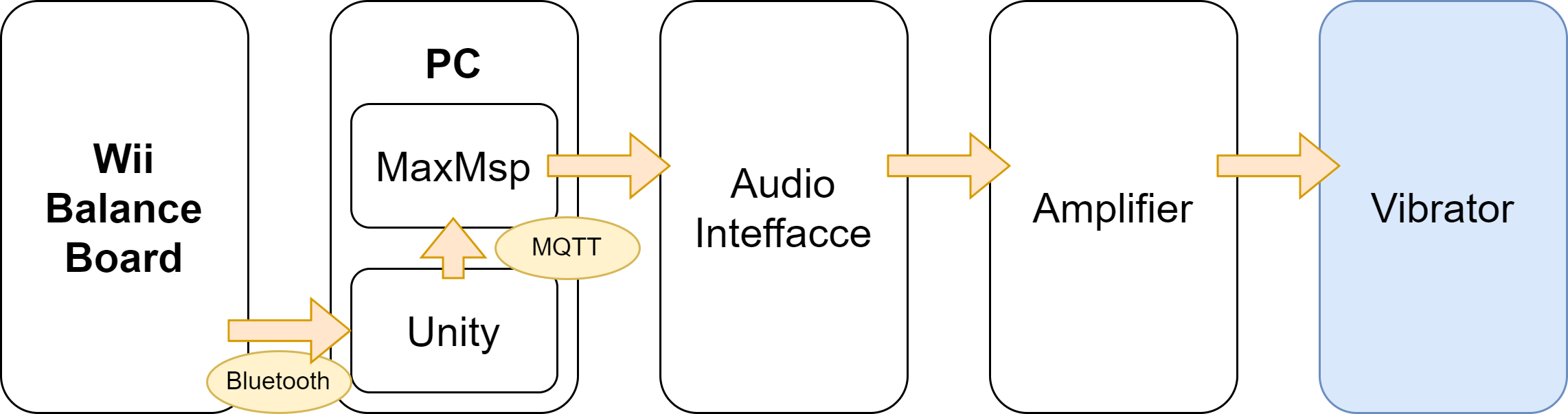}
 \caption{System Flow}
 \label{flow}
 \vspace{-5mm}
\end{center}
\end{figure*}

\begin{figure}[t]
\begin{center}
\includegraphics[width=\linewidth]{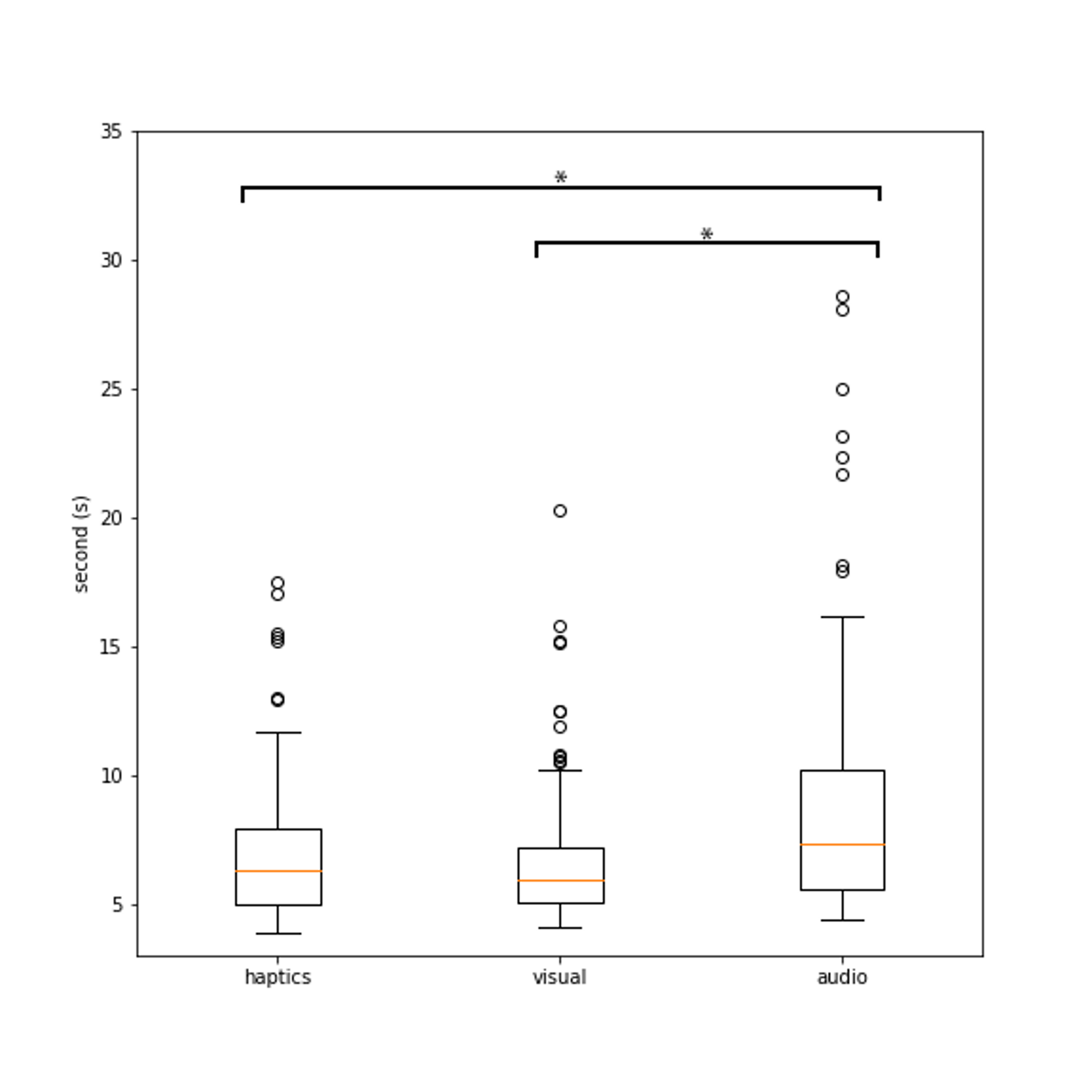}
\caption{\label{atof} Measured Time of Each Participants}
\end{center}
\end{figure}

\begin{figure}[t]
\begin{center}
\includegraphics[width=\linewidth]{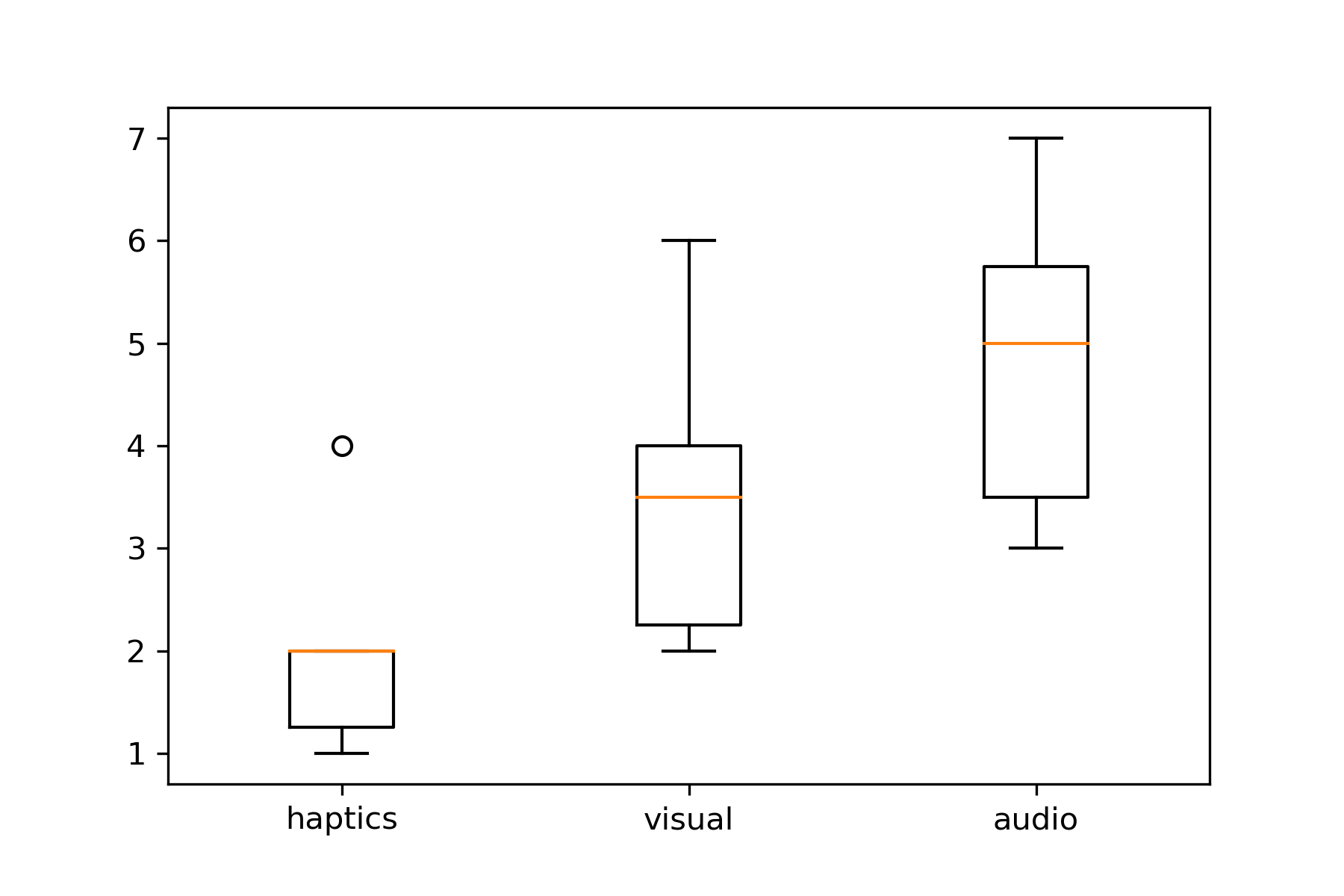}
\caption{\label{richard} Task difficulty for each feedback method}
\end{center}
\end{figure}

\begin{table}[t]
\begin{center}
\caption{\label{partinfo} Participants' information}
\label{tab:bpmsettings}
\begin{tabular}{c c}
\hline
Number & 6 males \\ 
Age & 23.0 $\pm$ 0.0 years old \\ \hline
\end{tabular}
\end{center}
\end{table}

\subsection{Experiment Overall}

Participants were tasked with moving their own CoP to specified locations, and the time taken for this movement was measured. Eight target points were set on the Wii Balance Board. These points were positioned 8 cm away from the center of the Wii Balance Board at 45-degree intervals (Fig.\ref{copmap}). An allowable region was defined within a radius of 3 cm from each set point, and the CoP was considered guided successfully if it remained within the allowable region for 3 seconds.
The target points were randomly presented to participants, with each point being presented three times. Participants experienced three different guidance patterns (haptic, visual, auditory) when guiding their CoP, as detailed later. A questionnaire survey was conducted after the experiments under all conditions. Participant information is presented in Table \ref{partinfo}.

\subsection{Implementaion}
The system flow implemented in this study is shown in Fig.\ref{flow}. The PC was connected to the Wii Balance Board using Bluetooth, and the CoP's position was continuously acquired. The sensor values received on Unity were sent to MaxMSP via MQTT communication. The feedback was then changed for each guidance method.

\subsection{Guidance Methods}
For the purpose of comparison with haptic guidance, we implemented guidance using visual and auditory feedback.

\begin{itemize}
\item Haptic Guidance: Participants were equipped with four vibration motors (Acouve Laboratory, Inc. Vibration Transducer Vp210) around their abdomen to receive haptic guidance. These motors were attached in the front, back, left, and right directions. The motors corresponding to the direction in which the participant's CoP needed to be guided would vibrate. During guidance, the motors vibrated at 100Hz, and this frequency was changed to 150Hz when the CoP entered the allowable region, providing feedback.
An example of the vibration pattern for haptic feedback is illustrated in Fig.\ref{vib}. In this figure, the participant's CoP is positioned in the lower left corner, while the target location is in the upper right corner. In this scenario, since the participant needed to move their CoP in the forward-right direction, the motors in the forward and right directions would vibrate.

\item Visual Guidance: Participants received guidance based on the position of points displayed on a screen. A square was shown along with a point positioned inside it. This point represented the direction in which the participant should move their CoP. In other words, participants aimed for the point to align with the center of the square. Feedback was provided by constantly displaying whether they were inside the allowable region.

\item Audio Guidance: Participants wore headphones and received auditory guidance through human voice. The voice prompts were pre-recorded. The direction in which to move the CoP was conveyed using clock positions (e.g., direction of 2 o'clock, 11 o'clock). When the CoP was within the allowable region, feedback was provided by playing the voice prompt "Answer."
\end{itemize}

\section{Result}

\subsection{Comparison of Required Time}
The required time and significance of each guidance pattern are shown in Fig.\ref{atof}.
The average time required for haptic feedback induction was $7.0 \pm 2.7 seconds$, for visual feedback induction was $6.6 \pm 2.5 seconds$, and for auditory feedback induction was $8.9 \pm 4.7 seconds$.
A Bonferroni-corrected multiple comparison test was used to investigate the statistical significance.
As a result of the tests, a significant difference was observed between haptic guidance and auditory guidance. A significant difference was also observed between visual guidance and auditory guidance.

\subsection{Questionnaire}

After the experiment, a questionnaire was conducted with the participants. Out of the six participants, four mentioned that haptic feedback was more understandable, while two stated that visual feedback was clearer.

The comments from those who mentioned that haptic feedback was more understandable were as follows:
\begin{itemize}
\item Q1: Because it's intuitive.
\item Q2: It's easier to understand the direction to move in based on the direction of vibration.
\item Q6: Visual feedback was unclear and insufficient. Auditory feedback became confusing when multiple were involved, and imagining a clock was unfamiliar. 
\end{itemize}

The comments from those who mentioned that visual feedback was more understandable were as follows:
\begin{itemize}
\item Q3: It's easiest to understand which direction (toward the target) to move in.
\item Q5: The direction to the answer was most intuitively understood, and adjusting the position was easy.
\end{itemize}
Furthermore, the survey results for task difficulty are shown in Fig.\ref{richard}. Participants were asked to rate the task difficulty for each feedback method on a scale of 1 to 7, where 1 indicated easy and 7 indicated difficult. The average difficulty for haptic feedback was $2.0 \pm 1.0$, for visual feedback was $3.5 \pm 1.4$, and for auditory feedback was $4.8 \pm 1.5$.

\section{Discussion ans Limitations}\label{SCM}
The results of the experiment revealed that inducing changes in CoP using haptic feedback is indeed feasible. When comparing the time taken for CoP induction between haptic and visual feedback as shown in Fig.\ref{atof}, no significant difference was observed. This suggests that haptic feedback is as effective as visual feedback in inducing CoP changes.

In the survey, many participants expressed that haptic feedback was the most intuitive and understandable. However, both haptic and visual feedback garnered opinions of being "intuitive and easy to understand." This underscores the importance of intuitive comprehension for participants. To achieve shorter induction times with haptic feedback, it is necessary to aim for even more intuitive feedback.

The task employed in this experiment was limited, involving moving one's CoP to a point 8 cm from the center of the Wii Balance Board. In the future, consideration should be given to experiments that simulate real-life movement scenarios.

\section{Conclusion}
We utilized vibration motors for haptic feedback to induce changes in participants' CoP. For comparison, we implemented induction using visual and auditory feedback and measured the time required for the induction to be completed. Although the induction through visual feedback had the shortest average time, the results of a multiple comparison test indicated no significant difference in the time required between induction through visual and haptic feedback. According to the survey, haptic feedback was found to have the lowest perceived level of difficulty.
In the future, our focus will be on inducing CoP in real-life movement scenarios.

\addtolength{\textheight}{-12cm}   % This command serves to balance the column lengths

%%%%%%%%%%%%%%%%%%%%%%%%%%%%%%%%%%%%%%%%%%%%%%%%%%%%%%%%%%%%%%%%%%%%%%%%%%%%%%%%

%%%%%%%%%%%%%%%%%%%%%%%%%%%%%%%%%%%%%%%%%%%%%%%%%%%%%%%%%%%%%%%%%%%%%%%%%%%%%%%%

%%%%%%%%%%%%%%%%%%%%%%%%%%%%%%%%%%%%%%%%%%%%%%%%%%%%%%%%%%%%%%%%%%%%%%%%%%%%%%%%

% \section*{ACKNOWLEDGMENT}
% This work was supported by JSPS KAKENHI Grant Number JP21H03485.

%%%%%%%%%%%%%%%%%%%%%%%%%%%%%%%%%%%%%%%%%%%%%%%%%%%%%%%%%%%%%%%%%%%%%%%%%%%%%%%%

\end{document}